\documentclass[journal]{./IEEEtran}
\ifCLASSINFOpdf
  \usepackage[pdftex]{graphicx}
  \DeclareGraphicsExtensions{.pdf,.jpeg,.png,.eps}
\else
\fi
\usepackage{cite}
\usepackage{graphicx}
\usepackage[cmex10]{amsmath}
\usepackage{algorithm}
\usepackage{algorithmic}
\usepackage{amssymb}
\usepackage{breqn}
\usepackage{array}
\usepackage[caption=false,font=normalsize,labelfont=sf,textfont=sf]{subfig}
\usepackage{url}
\usepackage{placeins}
\usepackage{multirow}
\usepackage{lipsum}
\usepackage{mathtools, cuted}
\usepackage{multicol, blindtext}
\usepackage{amsthm}

\newtheorem{theorem}{Theorem}
\usepackage{threeparttable}

\begin{document}
\title{Simultaneous Wireless Information and Power Transfer for Decode-and-Forward Multi-Hop Relay Systems in Energy-Constrained IoT Networks}
\author{\IEEEauthorblockN{Derek~Kwaku~Pobi~Asiedu,~\IEEEmembership{Student Member,~IEEE}, Hoon~Lee,~\IEEEmembership{Member,~IEEE}, and Kyoung-Jae~Lee,~\IEEEmembership{Member,~IEEE}}
\thanks{Manuscript received April 22, 2019; revised July 19, 2019; accepted August 19, 2019. This work was supported in part by the National Research Foundation of Korea (NRF) grant through Korea Government (MSIT) (NRF-2019R1A2C4070361), in part by the Institute for Information and Communications Technology Promotion (IITP) grant funded by Korea Government (MSIT) (No. 2018-0-00812, IoT wireless powered cognitive radio communications with user-centric distributed massive MIMO systems), and in part by the Korea Evaluation Institute of Industrial Technology (KEIT) grant funded by the Korea Government (MOTIE) (No. 10079984, Development of non-binding multimodal wireless power transfer technology for wearable device).}
\IEEEcompsocitemizethanks{A part of this work will be presented at the IEEE Vehicular Technology Conference Fall 2019 \cite{Asiedu19} as a conference version of this work.}
\IEEEcompsocitemizethanks{\IEEEcompsocthanksitem Derek~Kwaku~Pobi~Asiedu and Kyoung-Jae~Lee are with the Department of Electronics and Control Engineering, Hanbat National University, Daejeon 34158, South Korea.\protect

Hoon Lee is with the Department of Information and Communications Engineering, Pukyong National University, Busan, South Korea.\protect
 
Corresponding author: Kyoung-Jae Lee (kyoungjae@hanbat.ac.kr).}

\IEEEcompsocitemizethanks{Copyright (c) 2019 IEEE. Personal use of this material is permitted. However, permission to use this material for any other purposes must be obtained from the IEEE by sending a request to pubs-permissions@ieee.org.}
}

\maketitle

\begin{abstract}
This paper studies a multi-hop decode-and-forward (DF) simultaneous wireless information and power transfer (SWIPT) system where a source sends data to a destination with the aid of multi-hop relays which do not depend on an external energy source. To this end, we apply power splitting (PS) based SWIPT relaying protocol so that the relays can harvest energy from the received signals from the previous hop to reliably forward the information of the source to the destination. We aim to solve two optimization problems relevant to our system model. First, we minimize the transmit power at the source under the individual quality-of-service (QoS) threshold constraints of the relays and the destination nodes by optimizing PS ratios at the relays. The second is to maximize the minimum system achievable rate by optimizing the PS ratio at each relay. Based on convex optimization techniques, the globally optimal PS ratio solution is obtained in closed-form for both problems. By setting the QoS threshold constraint the same for each node for the source transmit power problem, we discovered that either the minimum source transmit power or the maximum system throughput can be found using the same approach. Numerical results demonstrate the superiority of the proposed optimal SWIPT PS design over conventional fixed PS ratio schemes.
\end{abstract}
%
\IEEEpeerreviewmaketitle
\section{Introduction}
\label{secintro}
\IEEEPARstart{T}HE Internet of Things (IoT) is the interconnection of network-enabled devices communicating with each other over the Internet ~\cite{Da14,Oteafy17,Deng18,Adhatarao18}. Hence, the purpose of IoT is to integrate the physical world and the virtual world to attain a self-sustaining system ~\cite{Da14,Oteafy17,Deng18,Adhatarao18}. To achieve this integration in the near future, research into the application of IoT for the creation of smart-cities, smart-homes, smart-energy, smart-transportation and many more is growing ~\cite{Da14,Oteafy17,Deng18,Adhatarao18}. In IoT, access and interaction between devices are seamless due to the use of unique identifiers, sensors and communication technologies ~\cite{Ciuonzo12,Ciuonzo14,Da14,Oteafy17,Deng18,Adhatarao18}. The evolution of IoT started from wireless sensor networks (WSN) and developed into different heterogeneous networks interacting with each other over the internet ~\cite{Ciuonzo12,Ciuonzo14,Rossi16,Deng18}. A wireless network of spatially distributed autonomous devices acting as sensors to monitor and record physical or environmental conditions (e.g.s humidity, pressure, temperature, power-line voltages, and vital body functions) is called a WSN ~\cite{Raghavendra06,Yick08,Ciuonzo12,Ciuonzo14,Rossi16,Yetgin17}. These sensor nodes may range from a few hundreds to thousands in a sensor network ~\cite{Raghavendra06,Yick08,Yetgin17}. Each sensor node is equipped with an antenna, microcontroller, interfacing electric circuit, and an energy source ~\cite{Raghavendra06,Yick08,Yetgin17}. 

While a WSN and other sensor networks operate in their own private networks, IoT integrates the various sensor networks into an IoT network by providing them access to the internet ~\cite{Adhatarao18}. For an IoT network, by the use of routing schemes, a gateway supports communication between the sensor nodes and a central system where actions are taken based on the information received from the sensor network ~\cite{Raghavendra06,Yick08,Ciuonzo12,Ciuonzo14,Rossi16,Yetgin17}. The gateway of a sensor network within an IoT network communicates with the central system over the internet. The routing schemes facilitates either direct or relayed communication between a gateway and a particular sensor node ~\cite{Raghavendra06,Yick08,Yetgin17}. Relays not only facilitate information forwarding, but may lead to an increase in a communication system's throughput ~\cite{Bai15,Simmons16,Guo16,Song17}. Relays, however, depend on their resources (i.e., power and computational components) to aid the processing and transfer of signals ~\cite{Bi15,Lu15,Guo16}. However, sensor nodes in IoT networks have limited resources, especially power ~\cite{Raghavendra06,Yick08,Yetgin17,Guo16,Adhatarao18}. Energy harvesting (EH) can be used to reduce the strain on the power resource of the relaying sensor nodes. By harvesting energy from a portion of the received radio frequency (RF) signal at the relay node, the harvested energy can be a source of power for information signal relaying ~\cite{Ulukus15,Ding15,Guo16}. 

The technique used for EH from RF signals is known as wireless power transfer (WPT) ~\cite{Zhou16,Song17,Mahama17}. WPT involves the wireless transfer of electrical energy from a power source to a load ~\cite{Mahama17}. EH is the process of scavenging energy from an external energy source ~\cite{Zhou16,Song17,Mahama17}. Typical traditional energy sources include hydro, wind, solar and wind, however, these sources of energy are less stable due unruly factors such as weather ~\cite{Mahama17}. Therefore, RF energy signals can serve as a more stable source of electrical energy for self-sustaining cellular systems ~\cite{Zhou16,Song17,Mahama17}. This implies that, EH from RF via WPT can be used in sensor network to aid sensor nodes with power constraints when operating as relays ~\cite{Guo16}. WPT implementation in cellular communication systems is accomplished using two main approaches, namely, wireless powered communication networks (WPCN) and simultaneous wireless information and power transfer (SWIPT) \cite{Tang18,Asiedu18,Mahama17,Lee16,Yin17,Lee18}. SWIPT involves the transmission of wireless information signal and wireless power signal concurrently ~\cite{Mukhlif18,Jameel17,Perera17}. Time switching (TS) and power splitting (PS) are the two main techniques by which SWIPT is implemented ~\cite{Mukhlif18,Jameel17,Perera17}. The successive transmission of wireless information signal and wireless power signal is the technique used to accomplish WPCN ~\cite{Ramezani17,Niyato16}. 

A few investigations on dual-hop SWIPT relay node system configurations are presented in ~\cite{Mahama17,Asiedu18,Do17} and ~\cite{Liu17}. Using outage probability and throughput as performance matrices, the research in ~\cite{Mahama17} optimizes the PS ratio for a single amplify-and-forward (AF) relay node facilitating communication between two nodes. In ~\cite{Mahama17}, it is assumed that each relay node's battery is charged with the energy it harvests from a portion of the received RF signal. In addition, all communicating nodes were equipped with a single antenna. The work in ~\cite{Mahama17} also included an extension into AF-SWIPT relay node selection. In ~\cite{Asiedu18}, the authors  considered a single antenna AF dual-hop configuration with multiple relay nodes. Both the power control factors and the PS ratios for each relay node were optimized with the objective of maximizing the achievable rate. The work presented in ~\cite{Do17} focused on a non-orthogonal multiple access (NOMA) system where a MIMO base station (BS) communicates with a single antenna cell-center user and a single antenna cell-edge user. The authors proposed three cooperative downlink (DL) transmission schemes based on hybrid SWIPT and BS antenna selection. The hybrid SWIPT powered relaying between the BS and cell-edge user is done by the cell-center user. The authors in ~\cite{Do17} acquired closed-form solutions for outage probability for their proposed schemes. The schemes were then compared to both orthogonal multiple access and non-NOMA systems in their simulation results. In addition, the authors presented a discussion on the diversity gains and complexity requirements of the various proposed schemes. A two-hop multi-antenna decode-and-forward (DF) SWIPT relay scenario is investigated in ~\cite{Liu17}. The PS ratio and power allocation at the DF-SWIPT relay node were optimized to maximize the end-to-end system throughput. The authors in ~\cite{Liu17} proposed an optimal clustering algorithm and a greedy clustering algorithm which grouped the multiple antennas of the DF-SWIPT relay node into information detection and energy harvesting antenna sets. 

SWIPT schemes in multi-hop relay systems are researched in ~\cite{Chen17}, ~\cite{Xu17}, ~\cite{Mao15}, and ~\cite{Liu19}. The work in ~\cite{Chen17} presents a novel multi-hop relay transmission strategy, where energy is harvested by the source and the relay nodes from the co-channel interference. The PS ratio for the source and AF-SWIPT multi-hop relay nodes were optimized based on a given outage probability threshold. In ~\cite{Chen17}, the authors identified the maximum number of AF-SWIPT multi-hop nodes that can support information transfer between a source node and a destination node. The research in ~\cite{Xu17} studied the coexistence of primary users (PUs) and secondary users (SUs). The PUs and SUs interacted in an energy harvesting cognitive radio network implemented using time division multiple access (TDMA) technique. The multi-hop relaying scenario in ~\cite{Xu17} occurs during data transmission between the SU nodes. The SUs harvest energy from the PU's RF signals. By employing an iterative algorithm, the end-to-end throughput of the SUs is maximized based on the time and power resource optimization  in ~\cite{Xu17}. A DF-SWIPT and an AF-SWIPT multi-hop relay configurations considering both the TS ratio and the PS ratio techniques are investigated in ~\cite{Mao15}. The authors in ~\cite{Mao15} aimed to find the maximum number of nodes which support communication between a source and a destination given a rate threshold constraint. A SWIPT AF multi-hop system with multiple-input-multiple-output (MIMO) relay nodes is investigated in ~\cite{Liu19}. In ~\cite{Liu19}, the achievable rate was maximized by optimizing the source and relay beamforming vectors, and the power resource of each node. 
 
In this paper, we investigate a DF sensor network, where a single antenna source node communicates with a single antenna destination node through multi-hop single antenna relay nodes. By adopting the PS ratio scheme, the multi-hop relay nodes operate in the SWIPT mode. Each relay node uses a portion of the RF signal it receives for EH, and it is saved in a supercapacitor. Unlike ~\cite{Chen17} and ~\cite{Xu17} which harvest energy from their co-channel interference and primary users, our model considers energy harvesting from only a portion of the RF signal received from the previous node. The harvested energy in the supercapacitor is used to both decode the rest of the RF signal, and forward the decoded information signal to the next node. An example in which our system model occurs is when the BS is requesting data from the destination wireless sensor node ~\cite{Raghavendra06,Sabor17,Djiroun17,Sandeep17}. The BS sends the request command to the destination wireless sensor via routing through the DF wireless sensor relay nodes. Another application of our system model involves device-to-device (D2D) communication between nodes facilitated by energy harvesting. Instead of relaying information, each node harvests energy from the RF signal it received from a previous node to power its own information signal transmission to the next node. Unlike the work in ~\cite{Mao15}, where the maximum number of hops is determined based on an available source power and throughput constraint, we solve the DF system throughput maximization and the source power minimization problems. In addition, we found a closed-form solution for determining the maximum number of DF-SWIPT relay nodes which can support communication between the source and the destination nodes given an SNR threshold constraint and an available power source value. Our close-form solution differs from the approach presented in ~\cite{Chen17} and ~\cite{Mao15}, which uses an algorithm to determine the maximum number of relay nodes. The main contributions of this paper are as follows.

\begin{itemize}
\item First, we seek to identify the minimum amount of source transmit power which supports communication between the source node and destination node via the DF-SWIPT multi-hop relay nodes. By tackling this problem, we reduce the strain (i.e., the depletion of the source power resource) on the source power during routing of information in the IoT network. From our source power minimization problem, we derive closed-form solutions for determining the minimum required transmit power at the source and the PS ratios for the DF-SWIPT relay nodes. 
\item We further consider the maximization of the minimum system achievable throughput also based on an already specified number of DF-SWIPT relay nodes and a source transmit power constraint. The maximization of the minimum system achievable rate is to aid in improving the system rate with different individual quality-of-service (QoS) constraint (i.e., the SNR thresholds) within the IoT network. We then derive closed-form solutions for the optimal PS ratios for each DF-SWIPT relay node, and the optimum system achievable throughput.
\item Using the optimal solutions for the source transmit power minimization problem and the optimal system achievable throughput problem, we show in this work that with a general SNR threshold constraint for all nodes within the IoT network, our two optimization solutions become equivalent. We treat equivalency of our solution due to the general SNR threshold consideration as a special scenario. 
\item From the special case, we present a closed-form solution to determine the number of relay nodes which support communication between the source node and destination node for a homogeneous sensor network. This closed-form solution can be used in a routing algorithm for our proposed system model. The prediction of the DF-SWIPT number can aid in resolving coverage, connectivity and routing issues in the IoT sensor networks ~\cite{Yetgin17,Mhatre04}.
\item Using the closed-form optimal PS ratio solution, we propose a centralized and a distributed method by which the PS ratio of each node can be determined in a real-world sensor network. In addition, we compare our two PS ratio determination methods in terms of complexity.
\item Since in practical systems channel estimation may not be perfect, within the simulation section of this paper we discuss the effects of imperfect channel estimation on the performance of our discussed optimization problems. 
\end{itemize}
From the remarks of authors in ~\cite{Yick08} and ~\cite{Yetgin17} concerning major issues on sensor networks, we can state the following advantages of implementing our proposed system model in a sensor network\footnote{The current issues as stated in ~\cite{Yick08} and ~\cite{Yetgin17} concerning sensor networks are with the limitations on the node powers, the computational power of each sensor node, the storage capacity of each sensor node, and the QoS requirement for the system.}. The application of our proposed sensor network scenario and optimization solutions presented in this work can be implemented in large or small sensor node network types. This is due to the low computational complexity of the closed-form solutions presented in this paper which is important considerations for sensor networks. Another advantage of this work is the possibility of lengthening network lifetime for a sensor network using energy harvesting. We also covered QoS requirement constraint (i.e. throughput), and resource limitation constraint (i.e. transmit power, computational and data storage capabilities) in this work. To the best of our knowledge, this is the first work to analyze both the source transmit power minimization problem and the optimization of the system achievable throughput problem for a DF-SWIPT multi-hop relay system model. We compare the optimal closed-form solutions to the fixed PS ratio scheme in our simulation results. The results affirmed that the optimal scheme outperformed the suboptimal scheme in terms of the minimum source power and the system achievable throughput. 

The rest of the paper is organized as follows: Section ~\ref{secsystem_model} presents the system model and optimization problem statement. Closed-form solutions for the optimization problems are discussed in Section ~\ref{secoptimization}. Section ~\ref{secsimulation} discusses our numerical results, and conclusions are drawn in Section ~\ref{secconclusion}.

\emph{Notations}: $n \sim \mathcal{CN}(0,\delta^{2})$ denotes a circularly symmetric complex Gaussian random variable, $n$, with zero mean and a variance of $\delta^{2}$. $\mathop{\mathbb{E}}_{X}[f(X)]$ is the expectation operation over random variable $X$. $f(X)$ and $R(X)$ represent a general function and a rate function respectively which are dependent on the variable $X$.  
\section{System Model and Problem Formulation}
\label{secsystem_model}
\begin{figure}[t!]
\begin{center}
\includegraphics[width=3.3in]{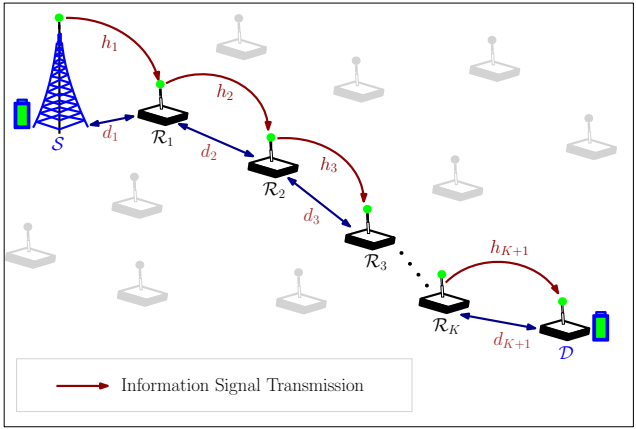}
\end{center}
\caption{Multi-hop DF relay systems with SWIPT architecture.}
\label{figSystems}
\end{figure}

A sensor network consisting of a source, $K$ multi-hop DF-SWIPT relay nodes, and a destination is investigated in this paper as shown in Fig. ~\ref{figSystems}. Each node is equipped with a single antenna. The source is a BS which may consist of the data gateway and the external/central systems of the sensor network ~\cite{Raghavendra06,Sabor17,Ciuonzo14,Rossi16,Djiroun17,Sandeep17}. The multi-hop DF-SWIPT relay nodes are the wireless sensor nodes found in the mesh network. This is due to the wireless sensor nodes being able to act as repeaters and relays ~\cite{Raghavendra06,Sabor17,Djiroun17,Sandeep17}. The destination node is also a wireless sensor node within the network ~\cite{Raghavendra06,Sabor17,Djiroun17,Sandeep17}. To reduce the strain on the relay node's resource (i.e., the power resource) during the relaying process, the relay nodes operate in the SWIPT mode. During the SWIPT mode, each relay utilizes their EH interfacing electrical circuit and a supercapacitor to facilitate the EH process from a portion of the RF signal it receives. Each relay node then facilitates the information decoding and forwarding process of the rest of the RF signal with the harvested energy.

\begin{figure}[t!]
\begin{center}
\includegraphics[width=3.4in]{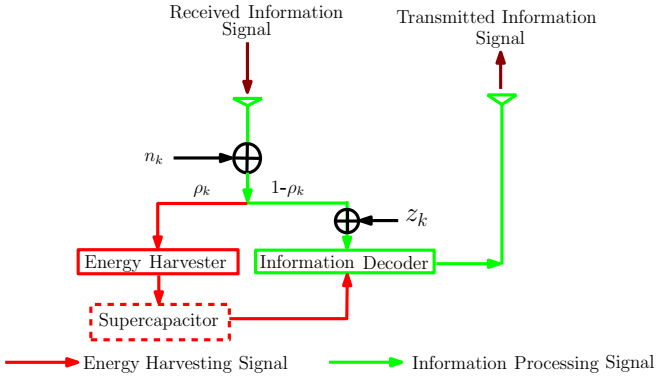}
\end{center}
\caption{Multi-hop DF relay node SWIPT architecture.}
\label{figSWIPTarch}
\end{figure}

The SWIPT architecture of our DF relay nodes in the sensor network system model is presented in Fig. ~\ref{figSWIPTarch}. The source, $\mathcal{S}$, and the destination, $\mathcal{D}$, already possess their own energy source for communication. Since each DF-SWIPT relay node uses a supercapacitor, each relay node dissipates all its harvested energy for information decoding (ID) and retransmission. Each DF-SWIPT relay node undergoes EH via the PS technique and operates in the half-duplex mode. We assume that the source node knows the channel state information (CSI) for all nodes communicating, while each DF-SWIPT relay node and destination node have knowledge of only the CSI for their communicating channels\footnote{The CSI for the sensor network can be acquired during the training phase for channel gain estimation when sensor nodes send pilots to each other and the BS (i.e., gateway and central system unit) ~\cite{Zhao04,Taricco12,Wang18}.}. It is assumed that, there is no direct link between the source and the destination nodes. Also, we assume that there is no direct link between the relay nodes. For example, from Fig. ~\ref{figSystems}, no direct link exists between $\mathcal{R}_{1}$ and $\mathcal{R}_{3}$. The assumption of no direct link holds for the worst case scenario, that is, when the distance between nodes is large. This assumption can be justified by virtue of routing from using a routing algorithm in the sensor network ~\cite{Liu19,Qiao13,Wu18}. The detailed operation of the considered system model will now be presented.

The received RF signal at node $k$ from the previous node is given as
\begin{equation}
y_{k}=\sqrt{E_{k-1}}h_{k}\hat{x}_{k}+n_{k}, \text{ }k=1,\ldots,K+1,
\end{equation}
where $h_{k}$ is the channel coefficient between the current node and the previous node, and $n_{k}\sim \mathcal{CN}(0,\delta^{2}_{k})$ represents the antenna noise at the current node. $\hat{x}_{k}$ stands for information signal from the preceding node. $K+1$ denotes the total number of subsequent nodes after the source node (i.e. the total number of DF-SWIPT relay nodes and the destination node). The channel $h_{k}$ is modeled as $h_{k}=\sqrt{\xi_{k}}\tilde{h}_{k}$, where $\xi_{k}$ is the large-scale fading coefficient and $\tilde{h}_{k}\sim \mathcal{CN}(0,1)$ represents the small-scale fading component with Rayleigh distribution ~\cite{Zhang13,Lee17}. The large-scale fading coefficient is modeled as $\xi_{k}=C_{k}\Big(\frac{d_{k}}{d_{0}}\Big)^{-\alpha_{k}}$, where $C_{k}$ is the constant attenuation for a reference distance $d_{0}$ ~\cite{Zhang13,Lee17}. $\alpha_{k}$ and $d_{k}$ indicate the pathloss exponent, and the distance between the transmit and receive nodes respectively ~\cite{Zhang13,Lee17}.

Next, the received RF signal at node $k$ is then split into two based on the PS ratio, $\rho_{k}$, for EH and ID. The EH and ID signals at node $k$ are respectively written as
\begin{equation}
\label{eq1}
y^{EH}_{k}=\sqrt{\rho_{k}}\Big(\sqrt{E_{k-1}}h_{k}\hat{x}_{k}+n_{k}\Big),
\end{equation}
and
\begin{equation}
\label{eq2}
y^{ID}_{k}=\sqrt{1-\rho_{k}}\Big(\sqrt{E_{k-1}}h_{k}\hat{x}_{k}+n_{k}\Big)+z_{k},
\end{equation}
where $z_{k}\sim \mathcal{CN}(0,\sigma^{2}_{k})$ is the additional noise introduced by the ID circuitry. 

From (~\ref{eq1}), the harvested energy, $E_{k}$, at node $k$ is expressed as
\begin{equation}
\begin{aligned}
\label{eq3}
E_{k}&\quad=\beta_{k}\mathop{\mathbb{E}}_{\hat{x}_{k},n_{k}}\Big[\vert y^{EH}_{k}\vert^{2}\Big]\backsimeq\beta_{k}\rho_{k}E_{k-1}\vert h_{k}\vert^{2}\\&\quad=E_{0}\Gamma_{k}\prod^{k}_{j=1}\rho_{j}, \text{ }k=1,\ldots,K+1,
\end{aligned}
\end{equation}
where $\Gamma_{k}\triangleq\prod^{k}_{j=1}\beta_{j}\vert h_{j}\vert^{2}$, $0<\beta_{k}\leq 1$ is the energy conversion efficiency of the $k$th node, and $E_{0}$ is the source transmit power. Also, from (~\ref{eq2}), the achievable rate, $R_{k}$, at node $k$ becomes
\begin{equation}
\begin{aligned}
\label{eq4}
R_{k}&\quad=\log_{2}\Bigg(1+\frac{E_{k-1}\vert h_{k} \vert^2(1-\rho_{k})}{(1-\rho_{k})\delta^{2}_{k}+\sigma^{2}_{k}}\Bigg)\\&\quad\backsimeq\log_{2}\Bigg(1+\frac{E_{k-1}\vert h_{k} \vert^2}{\sigma^{2}_{k}}(1-\rho_{k})\Bigg)\\&\quad=\log_{2}\Bigg(1+\frac{E_{0}\Gamma_{k}}{\sigma^{2}_{k}\beta_{k}}\prod^{k-1}_{j=1}\rho_{j}(1-\rho_{k})\Bigg),\\&\quad \text{ }k=1,\ldots,K+1,
\end{aligned}
\end{equation}
where (~\ref{eq4}) results from $\delta^{2}_{k}\ll \sigma^{2}_{k}$, that is, the antenna noise is negligible compared to the ID circuit noise power ~\cite{Liu13,Lee018,Zhang13}. 

In this paper, we consider two different optimization methods for the proposed multi-hop DF-SWIPT networks by jointly optimizing the PS ratio $\{\rho_{k}\}^{K}_{k=1}$ at the relay nodes. First, we aim to minimize the source transmit power, $E_{0}$, under the individual SNR constraint and PS ratio for each of multi-hop links as\footnote{To further emphasize the importance of solving the source power minimization problem, we will shortly present two possible scenarios in which our source power minimization problem is applicable as well as the DF multi-hop relay systems. The first deals with the system model where several node pairs perform SWIPT D2D communication with neighboring nodes (i.e., $\mathcal{R}_{k-1}\text{-to-}\mathcal{R}_{k}$ and $\mathcal{R}_{k}\text{-to-}\mathcal{R}_{k+1}$, $k=1,\ldots,K$ using Fig. ~\ref{figSystems} as a reference). Here, within the $\mathcal{R}_{k-1}\text{-to-}\mathcal{R}_{k}$ D2D nodes, $\mathcal{R}_{k}$ uses SWIPT PS ratio technique to decode and harvest energy from the RF signal it receives from $\mathcal{R}_{k-1}$. $\mathcal{R}_{k}$ then uses its harvested energy to forward its own information signal to $\mathcal{R}_{k+1}$ in the preceding D2D communication (i.e., $\mathcal{R}_{k}\text{-to-}\mathcal{R}_{k+1}$). The second application involves a SWIPT PS ratio D2D communication between the BS and $\mathcal{R}_{1}$ node. $\mathcal{R}_{1}$ then transfer energy to $\mathcal{R}_{2}$ to recharge its battery. The D2D WET occurs between $\mathcal{R}_{k}\text{-to-}\mathcal{R}_{k+1}$ for all subsequent nodes after $\mathcal{R}_{1}$ to recharge their batteries.}
\begin{equation}
\label{eq7}
\begin{aligned}
& \underset{E_{0},\{\rho_{k}\}^{K}_{k=1}}{\text{min}}\text{ }E_{0}\\
& \text{subject to} 
\begin{aligned}  
& & \frac{E_{0}\Gamma_{k}}{\sigma^{2}_{k}\beta_{k}}\prod^{k-1}_{j=1}\rho_{j}(1-\rho_{k})\geq \bar{\gamma}_{k}, \text{ }k=1,\ldots,K+1,
\end{aligned}\\
& \text{ } 
\begin{aligned}  
& & & & & & & & & & 0 \leq \rho_{k}\leq 1, \text{ }k=1,\ldots,K+1,
\end{aligned}\\
&\text{ } 
\begin{aligned}  
& & & & & & & & & & E_{0} \geq 0,
\end{aligned}
\end{aligned}
\end{equation}
where $\bar{\gamma}_{k}$ is the SNR threshold constraint of node $k$. 

Next, we consider the maximization of the overall DF system rate which can be formulated as
\begin{equation}
\label{eq5}
\begin{aligned}
& \underset{\{\rho_{k}\}^{K}_{k=1}}{\text{max}}\text{ }\text{ }
R\\
& \text{subject to} 
\begin{aligned}  
& & &0 \leq \rho_{k}\leq 1, \text{ }k=1,\ldots,K+1,
\end{aligned}
\end{aligned}
\end{equation}
where $R=\underset{\{1\leq k\leq K+1\}}{\text{min}}\text{ }\text{ }R_{k}$, and $R_{k}$ is defined in equation (~\ref{eq4}). With problem (~\ref{eq5}), the source transmit power is fixed. It is not considered as a constraint because we assume the source will transmit the RF signal with full power. 

Problem (~\ref{eq7}) is always feasible for any given set of the SNR constraints $\bar{\gamma}_{k}$ since the initial energy $E_{0}$ is included as an optimization variable in (~\ref{eq7}). For instance, it is obvious that the problem in (~\ref{eq7}) is feasible when $\gamma_{k}=0, \text{ }\forall k$. Also, even though some $\bar{\gamma}_{k}$ become large, we can always find a feasible PS ratio $\rho_{k}\in [0,1]$ by setting $E_{0}$ to a sufficiently large number. In addition, it is not difficult to show that the feasibility of the problem in (~\ref{eq5}) is always guaranteed since no constraints on the system throughput $R$ is included. Hence, both (~\ref{eq7}) and (~\ref{eq5}) are always feasible in practice. In the next section, we provide the globally optimal $\{\rho^{\star}_{k}\}^{K}_{k=1}$, $E^{\star}_{0}$ and $R^{\star}$ solutions for both problems. 

\section{Multi-Hop Relay Joint Optimal Design}
\label{secoptimization}
In this section, we solve the source transmit power minimization problem and the minimum system rate maximization problems in subsections ~\ref{subsecpowermin} and ~\ref{subsecrateminmax} respectively. We also discuss the physical (i.e., real-world) implementation of our optimal solutions and its influence on power constraint, computational constraint and QoS constraint in subsection ~\ref{subsecsecapplic}.
\subsection{Source Transmit Power Minimization}
\label{subsecpowermin}
In this subsection, the optimum values for the source transmit power and the PS ratio are presented in the Theorem ~\ref{thmoptimal_min}. We then provide a few remarks based on the optimal solutions for the source transmit power and the PS ratio of each DF-SWIPT relay node. 
\begin{theorem} \label{thmoptimal_min}
For the joint optimization problem given in (~\ref{eq7}), the optimal source transmit power, $E^{\star}_{0}$, is deduced as $$E^{\star}_{0}=\sum^{K+1}_{k=1}\frac{\bar{\gamma}_{k}\sigma^{2}_{k}\beta_{k}}{\Gamma_{k}},$$ with the optimal PS ratio, $\rho^{\star}_{k}$, at node $k$ defined as
$$\rho^{\star}_{k}=\begin{cases}
1-\frac{1}{\prod^{k-1}_{j=1}\rho_{j}}\frac{\frac{\bar{\gamma}_{k}\beta_{k}}{\Gamma_{k}}}{\sum^{K+1}_{j=1}\frac{\bar{\gamma}_{j}\beta_{j}}{\Gamma_{j}}},\text{ }\text{ }k=1,\ldots,K \\
0,\text{ }\text{ }\text{  }\text{ }\text{ }\text{ }\text{ }\text{ }\text{ }\text{ }\text{ }\text{ }\text{ }\text{ }\text{ }\text{ }\text{ }\text{ }\text{ }\text{ }\text{ }\text{ }k=K+1.
\end{cases}$$
\end{theorem}

\textit{Proof:} See Appendix ~\ref{app1}.

From Theorem ~\ref{thmoptimal_min}, we can infer that $E^{\star}_{0}$ is dependent on $\beta_{k}$, $\sigma_{k}$, $\Gamma_{k}$, and $\bar{\gamma}_{k}$, and independent of $\rho_{k}$. This implies that, the optimal source power can be determined at the source node without knowledge of each relay node's PS ratio. 

Secondly, we can further state that, increasing the number of DF-SWIPT relay nodes, increases the minimum source transmit power needed to support communication. Using an example to illustrate this deduction, for simplicity, we assume $\beta_{k}$, $\vert h_{k}\vert^{2}$, $\bar{\gamma}_{k}$ and $\sigma^{2}_{k}$ are the same for all DF-SWIPT relay nodes, that is, all the relays have similar properties\footnote{Similar properties here mean all the relays have the same channel structure, equal distances between each node, energy harvesting efficiency, and noise variance. This implies that, $\beta_{K+1}=1$, $\beta_{1}=\beta_{2}=\ldots=\beta_{K}=\beta$, $\vert h_{1}\vert^{2}=\vert h_{2}\vert^{2}=\ldots=\vert h_{K+1}\vert^{2}=\vert h \vert^{2}$, $\bar{\gamma}_{1}=\bar{\gamma}_{2}=\ldots=\bar{\gamma}_{K+1}=\bar{\gamma}$, and $\sigma^{2}_{1}=\sigma^{2}_{2}=\ldots=\sigma^{2}_{K+1}=\sigma^{2}$. Therefore, at the $k$th RF signal hop, we have $\Gamma_{k}$ becoming $\Gamma^{k}$ because $\Gamma_{1}=\beta\vert h \vert^{2}=\Gamma$, $\Gamma_{2}=\beta^2\vert h \vert^{4}=\Gamma^{2}$, and so on.}. $E^{\star}_{0}$ becomes $E^{\star}_{0}\thickapprox \sum^{K+1}_{k=1}\frac{\bar{\gamma}\sigma^{2}\beta}{\Gamma^{k}}$. As the number of nodes after the source node increases, $\Gamma^{k}$ reduces further below $1$, hence, leading $E^{\star}_{0}$ also increases. 

From the inspection of $\rho^{\star}_{k}$ in Theorem ~\ref{thmoptimal_min}, it can be established that $\rho^{\star}_{k}$ is dependent on the product of all previous nodes' PS ratio $\rho_{j}$ (i.e., $j=1,\ldots,k-1$) and independent of the optimal minimum source power. This implies that, the current $\rho^{\star}_{k}$ can not be determined without knowledge of the previous relay nodes' PS ratios. Since $\rho_k \propto \frac{1}{\prod^{k-1}_{j=1}\rho_{j}}$, this means the current DF-SWIPT relay node harvests less energy from the RF signal it receives from the previous node. Therefore, more portion of the received signal will be dedicated to ID at the current DF-SWIPT node. 
\subsection{Achievable System Rate Optimization}
\label{subsecrateminmax}
In this subsection, we will first reformulate the system rate optimization problem. From the reformulated problem, propose a theorem for finding the optimum system rate and the PS ratio of each relay node. Now, the system rate optimization problem for the DF-SWIPT relaying protocol is formulated as
\begin{equation}
\label{eq20}
\begin{aligned}
& \underset{\{\rho_{k}\}^{K}_{k=1}}{\text{max}}\text{ }\underset{\{1\leq k\leq K+1\}}{\text{min}}\text{ }
R_{k}\\
& \text{subject to} 
\begin{aligned}  
& & &0 \leq \rho_{k}\leq 1, \text{ }k=1,\ldots,K+1, 
\end{aligned}
\end{aligned}
\end{equation}
where $R_{k}=\log_{2}(1+\gamma_{k})$, and $\gamma_{k}$ represents the received SNR at node $k$ represented in equation (~\ref{eq4}). As $\gamma_{k}$ increases or decreases, the rate $R_{k}$ also increases and decreases. Hence, the system rate maximization problem (~\ref{eq20}) can be rewritten as   
\begin{equation}
\label{eq21}
\begin{aligned}
& R^{\star}=
\log_{2}\Bigg(1+\underset{\{\rho_{k}\}^{K}_{k=1}}{\text{max}}\text{ }\underset{\{1\leq k\leq K+1\}}{\text{min}}\text{ }\gamma_{k}\Bigg)\\
& \text{subject to} 
\begin{aligned}  
& & &0 \leq \rho_{k}\leq 1, \text{ }k=1,\ldots,K+1.
\end{aligned}
\end{aligned}
\end{equation}
By introducing a new system SNR constraint variable, $\hat{\gamma}$ (i.e., $\hat{\gamma}\leq \text{min}\text{ }\{\gamma_{k}\}^{K+1}_{k=1}$), the optimization problem (~\ref{eq21}) is redefined as 
\begin{equation}
\label{eq22}
\begin{aligned}
& \underset{\{\rho_{k}\}^{K}_{k=1},\hat{\gamma}}{\text{max}}\text{ }\hat{\gamma}\\
& \text{subject to} 
\begin{aligned}  
& & & \frac{E_{0}\Gamma_{k}}{\sigma^{2}_{k}\beta_{k}}\prod^{k-1}_{j=1}\rho_{j}(1-\rho_{k})\geq \hat{\gamma}, \text{ }k=1,\ldots,K+1,
\end{aligned}\\
& \text{ } 
\begin{aligned}  
& & & & & & & & & & 0 \leq \rho_{k}\leq 1, \text{ }k=1,\ldots,K+1,
\end{aligned}
\end{aligned}
\end{equation}
From problem (~\ref{eq22}), it can be deduced that $\hat{\gamma}$ is the minimum achievable SNR threshold of the system, hence, the solution to (~\ref{eq22}) should give us the optimal $\rho_{k}$ solution for which each hop SNR constraint is not less than $\hat{\gamma}$ ~\cite{Xu17,Liu17}. By solving (~\ref{eq22}), we obtain the following theorem.

\begin{theorem} \label{thmoptimal_minmax}
The optimal rate, $R^{\star}$, of the DF-SWIPT system is expressed as 
$$R^{\star}=\log_{2}\Bigg(1+\frac{E_{0}}{\sum^{K+1}_{k=1}\frac{\sigma^{2}_{k}\beta_{k}}{\Gamma_{k}}}\Bigg),$$ 
and the optimal PS ratio, $\rho^{\star}_{k}$, deduced as 
$$\rho^{\star}_{k}=\begin{cases}
1-\frac{1}{\prod^{k-1}_{j=1}\rho_{j}}\frac{\frac{\beta_{k}}{\Gamma_{k}}}{\sum^{K+1}_{j=1}\frac{\beta_{j}}{\Gamma_{j}}},\text{ }\text{ }\text{  } k=1,\ldots,K \\
0,\text{ }\text{ }\text{  }\text{ }\text{ }\text{ }\text{ }\text{ }\text{ }\text{ }\text{ }\text{ }\text{ }\text{ }\text{ }\text{ }\text{ }\text{ }\text{ }\text{ }\text{ }\text{ }\text{  }k=K+1.
\end{cases}$$
\end{theorem}

\textit{Proof:} See Appendix ~\ref{app2}.

From Theorem ~\ref{thmoptimal_minmax}, we make a few assertions. First, the current optimal PS ratio in (~\ref{eq34}) is dependent on the product of the PS ratios of all the preceding relay nodes. This means that the current node's PS ratio is smaller than the previous node's PS ratio. Hence, there may be more information decoded at the current node as compared to the previous node. Also, the optimal PS ratio is independent of the achievable SNR threshold. The achievable SNR threshold is also independent of the optimal PS ratio but depends on $\beta_{k}$, $\sigma_{k}$, $\Gamma_{k}$, and $E_{0}$. The optimal SNR threshold for the system reduces with increasing number of DF-SWIPT relay nodes between the source and the destination. 

\subsection{Implementation and Analysis}
\label{subsecsecapplic}
In this subsection, we discuss how the proposed protocols can be implemented. We discuss a special scenario where both the source power minimization and the minimum system rate maximization are equivalent. We also present a simple closed-form solution for the determination of the number of relay nodes need to support communication between the source and the destination. This solution is for the special scenario where the sensor network is homogeneous and the inter-node distance are equivalent (i.e., the distance between nodes are equal). The closed-form solution for determining the number of relay nodes can be used in a routing algorithm for our proposed system model. For the protocol implementation, we will delve into how the PS ratio can be calculated for each relay node based on the computation constraints of each sensor node. 

For the source power minimization problem, if each relay node has the same SNR threshold constraint (i.e., $\bar{\gamma}_{1}=\bar{\gamma}_{2}=\ldots=\bar{\gamma}_{K}=\bar{\gamma}_{K+1}$), then the PS ratio solution for both problems (~\ref{eq8}) and (~\ref{eq20}) are the same. That is, for both cases, the optimal PS ratio is expressed as
\begin{equation}
\label{eqgenrho}
\rho^{\star}_{k}=\begin{cases}
1-\frac{1}{\prod^{k-1}_{j=1}\rho_{j}}\frac{\frac{\beta_{k}}{\Gamma_{k}}}{\sum^{K+1}_{j=1}\frac{\beta_{j}}{\Gamma_{j}}},\text{ }\text{ }k=1,\ldots,K \\
0,\text{ }\text{ }\text{  }\text{ }\text{ }\text{ }\text{ }\text{ }\text{ }\text{ }\text{ }\text{ }\text{ }\text{ }\text{ }\text{ }\text{ }\text{ }\text{ }\text{ }\text{ }\text{ }k=K+1.
\end{cases}
\end{equation}
In addition, (~\ref{eq18}) and (~\ref{eq32}) become equivalent. This implies that, given either a QoS constraint (i.e., the system required SNR threshold) or a source power constraint (i.e., the minimum available source power), we can calculate either the minimum source power or the maximum system achievable rate, respectively, from the generalized equation,
\begin{equation}
\label{eqgeneraleq}
\frac{E_{0}}{\hat{\gamma}}=\sum^{K+1}_{k=1}\frac{\sigma^{2}_{k}\beta_{k}}{\Gamma_{k}}.
\end{equation} 

Knowing $E_{0}$ and $\hat{\gamma}$, we can estimate the number of relay nodes needed to support communication between the source node and the destination node from (~\ref{eqgeneraleq}). By manipulating the (~\ref{eqgeneraleq}), we can obtain the estimated maximum number of relay nodes as
\begin{equation}
\label{eqnodenumb}
K\thickapprox \begin{cases}
\frac{\ln\Big[1-\Big(\frac{E_{0}}{\bar{\gamma}\sigma^{2}\beta}+1\Big)\Big(1-\frac{1}{\Gamma}\Big)\Big]}{-\ln \Gamma}-1,\text{ }\text{ }\Gamma > 1 \\
\frac{\ln\Big[1+\Big(\frac{E_{0}}{\bar{\gamma}\sigma^{2}\beta}+1\Big)\Big(\frac{1}{\Gamma}-1\Big)\Big]}{-\ln \Gamma}-1,\text{ }\text{ }\Gamma <1,
\end{cases}
\end{equation}
where $\Gamma =0.5 \beta C\Big(\frac{d}{d_{0}}\Big)^{-\alpha}$, and the variables $C$, $d$, $\alpha$ are the inter-node attenuation constant, distance, and pathloss exponent. The $\Gamma$ and $K$ are predetermined at the source node and used in a routing algorithm. The detailed derivation of (~\ref{eqnodenumb}) is presented in Appendix ~\ref{app3}. 

The two possible methods for determining the PS ratio are by a centralized method and a distributed method. For the centralized method, since the source node knows all the CSI for all communicating channels in the network, it can calculate the PS ratio for each DF-SWIPT node. The source node calculates the PS ratio for each node using equations (~\ref{eq19}) and (~\ref{eq34}) for the source power minimization case and the system throughput maximization case, respectively. After calculating all the PS ratios, $\{\rho^{\star}_{k}\}^{K}_{k=1}$, it transmits the PS ratios with its information signal to the first DF-SWIPT node. With the centralized system, the relay node $k$ must transmit its decoded information along with $\{\rho^{\star}_{j}\}^{K}_{j=k+1}$ to the next relay node. However, with the distributed system, since each relay node knows its own CSI for the channels it communicates on, it can calculate the PS ratio for the next node. In the distributed system, the source node calculates the PS ratio of the first relay node as 
\begin{equation}
\rho_{1}=1-\tilde{\psi}_{1},
\end{equation}
where $\tilde{\psi}_{1}=\frac{1}{\vert h_{1} \vert^2\sum^{K+1}_{j=1}\frac{\beta_{j}}{\Gamma_{j}}}$. The source node then transmits its information signal, $\rho_{1}$ and $\tilde{\psi}_{1}$ to the first relay node. The $k$th relay node calculates the $k+1$ relay's PS ratio as 
\begin{equation}
\rho_{k+1}=1-\tilde{\psi}_{k+1},
\end{equation} 
where $\tilde{\psi}_{k+1}=\frac{1}{\rho_{k}\beta_{k}\vert h_{k+1} \vert^2}\tilde{\psi}_{k}$, and $k=1,\ldots,K-1$. The current $k$ DF-SWIPT node transmits its decoded information, $\rho_{k+1}$ and $\tilde{\psi}_{k+1}$ to the next $k+1$ DF-SWIPT node. The advantage of the centralized system is with the relay nodes not having any computational burden concerning the PS ratio calculation. However, the first few relay nodes would have a large amount of data bits to process and forward depending on the number of preceding relay nodes' PS ratios transmitted to it. The DF process may be affected if the relay nodes do not have enough memory (i.e., computational processing power). But, with the distributed system, each node receives fewer information bits to DF compared to the centralized system. The drawback of the distributed system is the need for the DF-SWIPT relay node to compute variables $\rho_{k+1}$ and $\tilde{\psi}_{k+1}$ before retransmission to the next node.

A comparison of the centralized and distributed methods is summarized in Table ~\ref{tabsummary}.  For the computational complexity comparison, let $\mathcal{O}(J)$ represent a single arithmetic operation. For the centralized system, the source node performs $\mathcal{O}((K+1)J)$ arithmetic operations, that is, it calculates the $E_{0}$ and $\{\rho_{k}\}^{K}_{k=1}$ values. The relay nodes do not perform any arithmetic operations in the centralized system. However, with the distributed method, the source node performs $\mathcal{O}(J)$ arithmetic operations to determine $E_{0}$, $\tilde{\psi}_{1}$ and $\rho_{1}$, while the $k-$th relay node performs $\mathcal{O}(J)$ arithmetic operations to determine $\tilde{\psi}_{k+1}$ and $\rho_{k+1}$. 

\begin{table*}
\centering
\caption{Comparison of PS Ratio Determination Methods}
\label{tabsummary}
\begin{tabular}{|c|c|c|c|c|}
\hline
PS ratio Scheme & \multicolumn{2}{c|}{Centralized Method} & \multicolumn{2}{c|}{Distributed Method} \\ \hline\hline
Communicating Node & Source & $k-$th Relay Node  & Source & $k-$th Relay Node\\  \hline
Computational Complexity & $\mathcal{O}((K+1)J)$ & --  & $\mathcal{O}(J)$ & $\mathcal{O}(J)$\\  \hline
Transmit Bits & $K (i_{0}\log_{2} K + B) + F $  & $(K-k) \{i_{0} \log_{2}(K-k) + 2B\} + F $ & $2B + F$ & $ 2B + F $ \\  \hline
CSI Requirement & Global CSIs & -- & Global CSIs & Local CSIs  \\ \hline
\end{tabular}
\end{table*}

Next, we discuss the number of bits needed to be transmitted at each node. First, we assume the actual transmitted information bit, real number bits (i.e. either $\rho_{k+1}$ or $\tilde{\psi}_{k+1}$ value), and relay index bits broadcast from the current node to the next node to be processed are defined as $F$, $B$, and $i_{0}$, respectively. With the centralized system, the source node transmits the information signal, the $K$ relay indexes and $K$ PS ratios to the first relay node as $K (i_{0}\log_{2} K + B) + F$ bits. Each $k-$th relay node then transmit its decoded information signal, the $K-k$ relay indexes and $K-k$ PS ratios to the next relay node. For the distributed method, each node including the source node transmits its information signal, and the next node's PS ratio and $\tilde{\psi}_{k+1}$ value to the next relay node as $2B + F$ bits.

\section{Simulation Results}
\label{secsimulation}
This section provides simulation results to demonstrate the system performance of multi-hop DF-SWIPT sensor network. Unless otherwise stated, the following parameters are utilized for the simulations: for the model channel model presented in paragraph three of Section ~\ref{secsystem_model}, for the large-scale fading component, the attenuation constant $C_{0}=-10$dB, the pathloss exponent $\alpha = 3$ and the inter-node distance is set as $d=d_{1}=d_{2}=\ldots=d_{K}=2$m. Please, note that by considering $d_{1}=d_{2}=\ldots=d_{K}=2$m, the distance between the source node and the destination node increases with increasing number of DF-SWIPT relay nodes. Hence, the total distance between the source node and destination node for $K$ DF-SWIPT relay nodes is $2\times (K+1)\times d$. The antenna noise variance $\sigma^{2}_{1}=\sigma^{2}_{2}=\ldots=\sigma^{2}_{K+1}=-80$dBm, and the energy conversion efficiency $\beta_{1}=\beta_{2}=\ldots=\beta_{K}=0.7$. A suboptimal naive scheme with a fixed PS ratio ($\rho_{k}=0.5$, $\forall k$) is adopted for comparison with the optimal DF-SWIPT scheme. The simulation results are obtained over $10^{4}$ random channel realizations. 
\begin{figure}[t!]
\begin{center}
\includegraphics[width=3.4in]{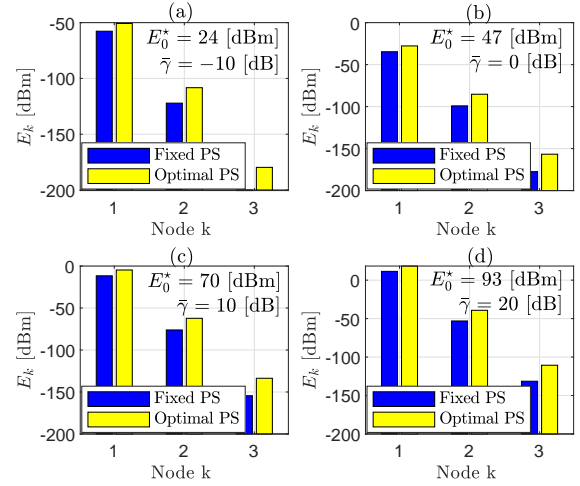}
\end{center}
\caption{Comparison of various schemes based on harvested power at each relay node, $K=3$ and $d_{k}=2$m.}
\label{figresults5}
\end{figure}
 
Fig. ~\ref{figresults5} shows a comparison of the energy harvested by the optimal DF-SWIFT scheme and suboptimal DF-SWIFT scheme at each node. The plot of each relay node's harvested energy is presented for $K=3$, a set of SNR threshold constraints (i.e. $\bar{\gamma}=-10$dB, $0$dB, $10$dB and $20$dB). For each SNR threshold in the figure, the optimal source transmit power is calculated for $K=3$, and used in determining how much energy is harvested at each node. From the figure, it can be seen that the fixed $\rho_{k}$ scheme harvests lesser energy at each node as compared to the optimal PS ratio. There is a reduction in the amount of harvested energy as the RF signal moves from one node to the other. This may lead to a subset of all nodes being able to harvest enough energy to support communication between the source and destination for the suboptimal scheme as shown in the first plot of Fig. ~\ref{figresults5}(a). Therefore, to achieve the same QoS, the source needs to transmit more power for the suboptimal scheme. Hence, the suboptimal scheme puts a higher strain on the source power. 
\begin{figure}[t!]
\begin{center}
\includegraphics[width=3.2in]{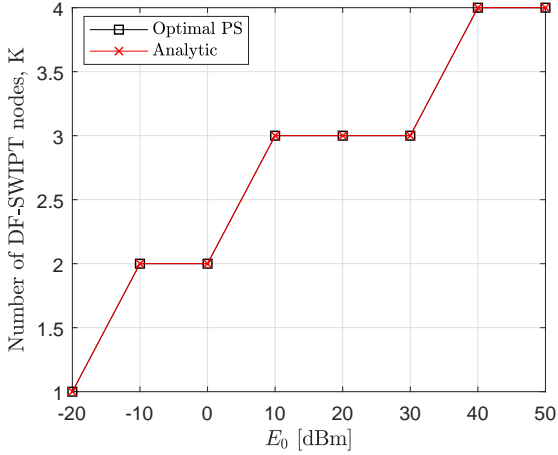}
\end{center}
\caption{The maximum number of DF-SWIPT nodes against increasing source transmit power, $\bar{\gamma}=20$dB and $d_{k}=2$m.}
\label{figresults11}
\end{figure}
\begin{figure}
\begin{center}
\includegraphics[width=3.2in]{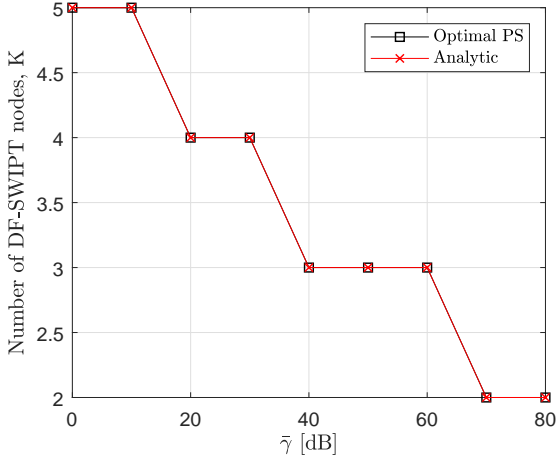}
\end{center}
\caption{The maximum number of DF-SWIPT nodes against increasing system required SNR, $E_{0,max}=50$dBm and $d_{k}=2$m.}
\label{figresults10}
\end{figure}
Figs. ~\ref{figresults11} and ~\ref{figresults10} show the implementation of (~\ref{eqnodenumb}) in determining the number of relay nodes. Fig. ~\ref{figresults11} shows the plot of the number of relay nodes against varying $E_{0}$, while, Fig. ~\ref{figresults10} is a plot on the number of relay nodes against varying $\bar{\gamma}$. From both plots, we can observe that the closed-form solution for determining the number of nodes gives an excellent approximation of the number of relay nodes.
\subsection{Transmit Power Minimization}
\label{subsecsulenergymin}
\begin{figure}[t!]
\begin{center}
\includegraphics[width=3.2in]{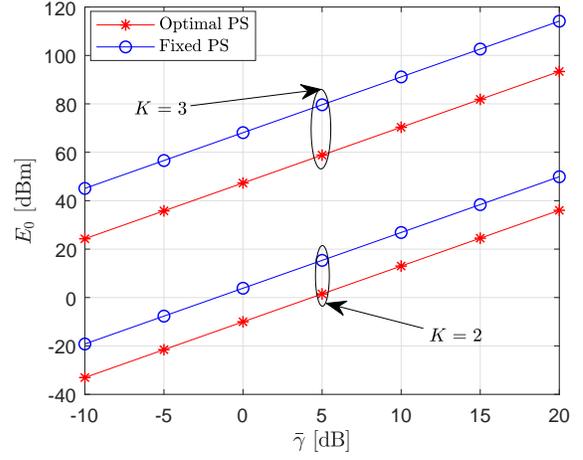}
\end{center}
\caption{Average minimum transmit power with respect to varying $\bar{\gamma}$ plot for $K=2$ and $3$, and $d_{k}=2$m.}
\label{figresults1}
\end{figure}
Fig. ~\ref{figresults1} shows a graph of the minimum source transmit power, $E_{0}$, against the SNR threshold constraint range, $\bar{\gamma}$, for different number of relay nodes, i.e., $K=2$ and $K=3$. The optimal PS scheme outperforms the fixed PS suboptimal scheme in terms of the $E_{0}$ needed to support the source to destination communication. The optimal scheme achieves a $15$dBm gain over the suboptimal scheme when $K=2$ and $20$dBm gain with $K=3$. An increase in the SNR threshold generally results in a corresponding increase in the minimum source power required for successful communication. Also, a rise in the number of relay nodes results in an increase in the minimum source power. This conclusion is consistent with the analysis of Theorem ~\ref{thmoptimal_min}. It is observed that there is a loss of about $60$dBm and $50$dBm of the source transmit power in the optimal and the suboptimal PS schemes, respectively, with just a single increase in the number of relay nodes. This increase in $E_{0}$ is due to the increase in the distance and the number of relay nodes between the source and the destination nodes.
\begin{figure}[t!]
\begin{center}
\includegraphics[width=3.2in]{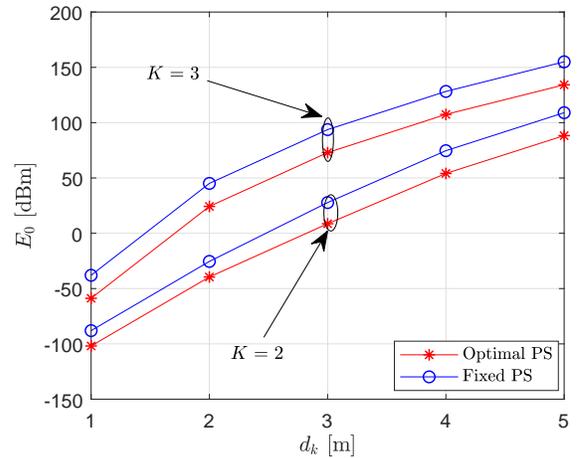}
\end{center}
\caption{Average minimum transmit power  for increasing distance between nodes with $K=2$ and $3$, and $\bar{\gamma}=-10$dB.}
\label{figresults3}
\end{figure}
 
Fig. ~\ref{figresults3} shows a plot of the minimum source power , $E_{0}$, against the inter-node distance, $d_{k}$. The average minimum amount of source transmit power needed to support the end-to-end communication in the sensor network rises with increasing $d_{k}$ for both the optimal and suboptimal PS schemes. There is a constant difference in the $E_{0}$ for the optimal and suboptimal for all values of $d_{k}$. A performance degradation of about $15$dBm and $25$dBm occurs between the optimal and suboptimal schemes for both $K=2$ and $K=3$, respectively as $d_{k}\geq 2$m. By increasing the number of relays, there is a significant rise in the $E_{0}$ needed to facilitate communication. This phenomenon is due to the increase in distance between the source and destination nodes which in-turn influences the signal attenuation over the increasing inter-node distance. 

\begin{figure}[t!]
\begin{center}
\includegraphics[width=3.2in]{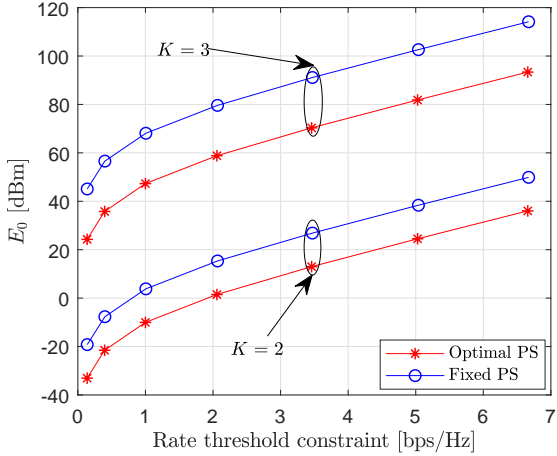}
\end{center}
\caption{Average minimum transmit power against rate threshold for $K=2$ and $3$.}
\label{figresults4}
\end{figure}
A plot of the average minimum $E_{0}$ against the system rate threshold constraint is presented in Fig. ~\ref{figresults4}. Similarly, by increasing the system rate threshold, the system's minimum source transmit power, $E_{0}$, requirement increases. From the system rate threshold of $1$bps/Hz upward, there is an improvement of about $15$dBm in terms of $E_{0}$ for the optimal PS scheme, and a $20$dBm improvement for the suboptimal PS scheme.

\begin{figure}[t!]
\begin{center}
\includegraphics[width=3.2in]{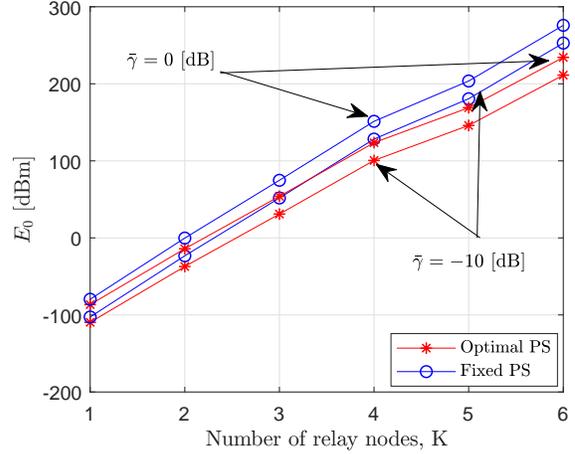}
\end{center}
\caption{Average minimum transmit power against the number of DF-SWIPT relay nodes at  $\bar{\gamma}=-10$ and $0$dB, and $d_{k}=2$m.}
\label{figresults2}
\end{figure}

This behavior is well appreciated in Fig. ~\ref{figresults2}, where we plot the minimum source power against the number of relay nodes. Here, we set $d_{k}=2$m. The distance between the source and destination node increases from $4$m to $14$m for $K=1$ and $K=6$, respectively. Increasing the DF-SWIPT nodes produces a constant increase in the minimum source transmit power needed to support the system QoS. 
\subsection{System Rate Maximization}
\label{subsecsulratemaxmin}
\begin{figure}[t!]
\begin{center}
\includegraphics[width=3.2in]{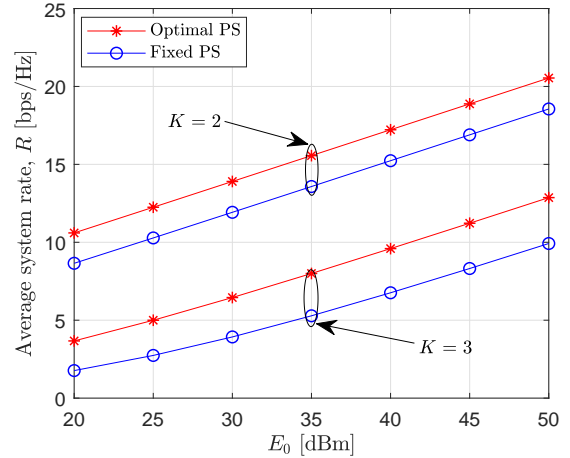}
\end{center}
\caption{Average achievable system rate against available source power, $K=2$ and $3$, and $d_{k}=2$m.}
\label{figresults7}
\end{figure}

Fig. ~\ref{figresults7} shows the impact of the source transmit power on the achievable rate for different numbers of relay nodes. The optimal PS scheme has better achievable rate as compared to the suboptimal PS scheme. The lesser the number of relay nodes, the higher the achievable rate. This is because the distance between the source and destination node also reduces. For the $E_{0}$ range of $30$dBm to $50$dBm, the optimal and suboptimal schemes have a performance gap of about $4$bps/Hz and $6$bps/Hz for $K=2$ and $K=3$, respectively. There is a performance improvement for both schemes of about $5$bps/Hz when relay node is reduced (i.e. from $K=3$ to $K=2$).

\begin{figure}[t!]
\begin{center}
\includegraphics[width=3.2in]{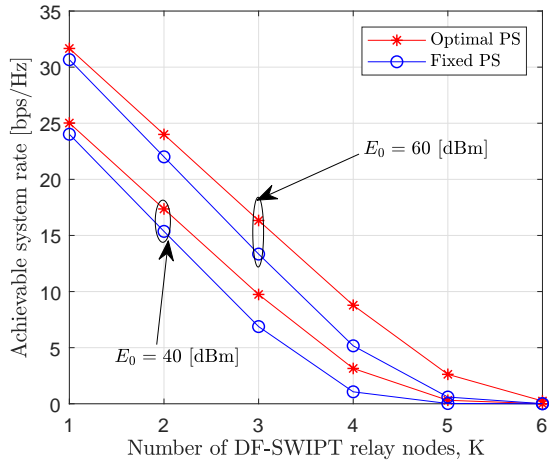}
\end{center}
\caption{Average achievable system rate against available source power, $E_{0}=30$ and $60$dBm, and $d_{k}=2$m.}
\label{figresults8}
\end{figure}

\begin{figure}[t!]
\begin{center}
\includegraphics[width=3.2in]{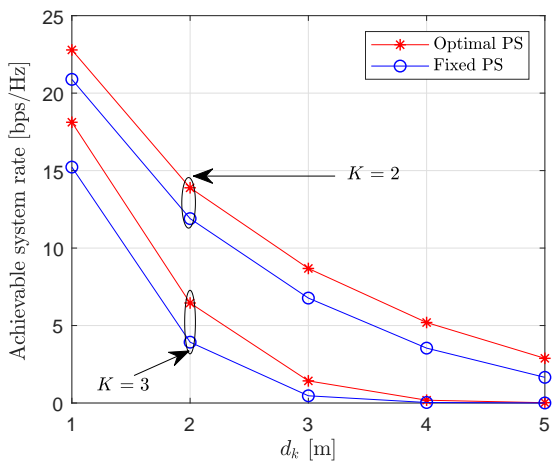}
\end{center}
\caption{Average achievable system rate against available source power, $E_{0}=30$dBm, and $K=2$ and $3$.}
\label{figresults9}
\end{figure}

To further appreciate the effect of how increasing the number of DF-SWIPT relay nodes has on the achievable rate, we consider Fig.~\ref{figresults8}. The plot shows the average achievable rate against the number of relays for $E_{0}=30$ and $60$dBm. From Fig. ~\ref{figresults8}, the achievable rate deteriorates with increasing number of relay nodes. For $K>5$, the achievable rate approaches zero for both schemes. A similar behavior can be seen in Fig. ~\ref{figresults9} where the achievable rate reduces with an increase in the inter-node distance. For a fixed number of relays, as the distance between the relays increases, the performance gap between the suboptimal and optimal schemes narrows. This is due to the widening of the distance between the source and destination node by a total distance of $d_{k}\times(K+1)$. 

\subsection{Imperfect Channel State Information}
\label{subsecimpcsicon}
We now consider imperfect channel state information (ICSI) for calculating the optimal PS ratio since channel estimation in practical systems may not be accurate. This inaccuracy may be due: (i) inherent delay occurring between the channel estimation and actual data transmission, and (ii) limited feedback in frequency division multiplexing or imperfect reciprocity in time division multiplexing ~\cite{Yoo06,Lee09,Xiang12}. The source node is assumed to have ICSI of each relay node ~\cite{Yoo06,Lee09,Xiang12}. With the ICSI, the channel estimation error is modeled as $\tilde{h}_{k}=\hat{h}_{k}+e_{k}$, where $\hat{h}_{k}\sim \mathcal{CN}(0,1-\sigma^{2}_{E})$ and $e_{k}\sim \mathcal{CN}(0,\sigma^{2}_{E})$ represent the estimated and the error channel coefficients, respectively. Here, $\sigma^{2}_{E}$ is the estimation error variance, which is assumed to be $0$, $0.1$, $0.2$, and $0.3$ for our simulation. 
\begin{figure}[t!]
\begin{center}
\includegraphics[width=3.35in]{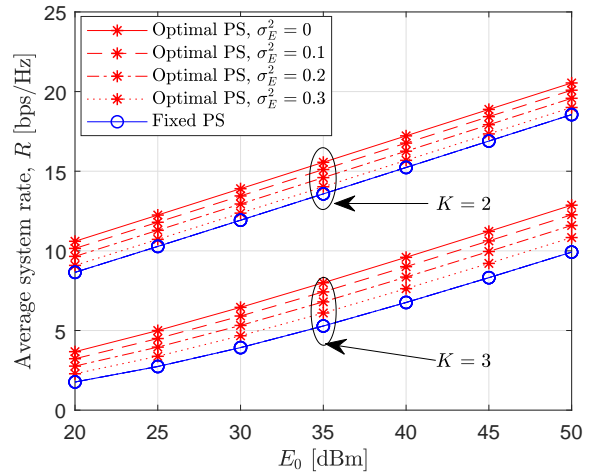}
\end{center}
\caption{The effect of estimated channel errors for $K=2$ and $3$, and $d_{k}=2$m.}
\label{figresultsimp1}
\end{figure}

The effect of ICSI on our achievable rate problem is presented in Fig. ~\ref{figresultsimp1}. As expected, when $\sigma^{2}_{E}$ increases, the maximum  achievable rate reduces. It can be observed from Fig. ~\ref{figresultsimp1} that the fixed PS ratio scheme remains unchanged for all $\sigma^{2}_{E}$ considered in the simulation, since it does not exploit the channel state information to compute the PS ratio. In addition, by reducing the estimation error variance, the curves in both plots approach their perfect CSI curves (i.e., $\sigma^{2}_{E}=0$) for optimal schemes. This result shows that the proposed rate maximization solution can be used in practical environment. 

\section{Conclusion}
\label{secconclusion}
This paper investigated a DF-SWIPT multi-hop relay configuration in an IoT sensor network scenario. The DF-SWIPT relay nodes facilitated communication between a source and a destination node. All the nodes possess a single antenna. Two optimization problems have been studied in this work. We first optimized the PS ratio with the aim of minimizing the source transmit power. We then optimized the PS ratio with the objective of maximizing the achievable rate. Closed-form solutions were found for the minimum source power, maximum system achievable rate and the PS ratio at each DF-SWIPT relay node. From our research, we found a single equation linking the source energy to the system achievable rate. We have also shown that the source power minimization problem is equivalent to the system throughput maximization. The optimal PS ratio scheme was compared to a fixed PS ratio suboptimal scheme. The optimal scheme outperformed the suboptimal scheme for both the source power minimization problem and the system throughput maximization problem in the simulation section.  

In this work, we consider the simple case of single antenna system for our initial research analysis and presentation of our DF-SWIPT proposed communication model. A multi-antenna systems as in ~\cite{Liu19} and imperfect channel estimation at each node can be considered as extension of this work. Unlike the works presented in ~\cite{Mao15} and ~\cite{Liu16} which considered both AF and DF relay configuration, we focus on only DF relay configuration in this paper. The AF relay configuration can be considered as an extension of this work. Another approach or extension which can be researched is the use of rechargeable battery instead of a supercapacitor as considered in ~\cite{Mahama17} and ~\cite{Lee016}. We also considered DF-SWIPT multi-hop relay system using the PS SWIPT technique in this work. Hence, the TS SWIPT technique can be considered as an extension of the  work presented in this paper.

\appendices 
\section{Proof of Theorem 1}
\label{app1}
First of all, problem (~\ref{eq7}) is non-convex with respect to both $E_{0}$ and $\{\rho_{k}\}^{K}_{k=1}$. By introducing four new variable definitions, we reformulate problem (~\ref{eq7}) into an equivalent convex problem as follows; 
\begin{equation}
\label{eq8}
\begin{aligned}
& \underset{Q,\{A_{k}\}^{K+1}_{k=1}}{\text{minimize}}\text{ }\frac{1}{Q}\\
& \text{subject to} 
\begin{aligned}  
& & & A_{k-1}-A_{k} \geq Q\frac{\sigma^{2}_{k}\beta_{k}\bar{\gamma}_{k}}{\Gamma_{k}}, \text{ }k=1,\ldots,K+1,
\end{aligned}\\
& \text{ } 
\begin{aligned}  
& & & & & & & & & & A_{k} \leq A_{k-1}, \text{ }k=1,\ldots,K+1,
\end{aligned}\\
&\text{ } 
\begin{aligned}  
& & & & & & & & & & Q \geq 0,
\end{aligned}
\end{aligned}
\end{equation}
where $A_{k}\triangleq\prod^{k}_{j=1}\rho_{j}$, $A_{0}\triangleq 1$, $A_{K+1}\triangleq 0$, and $Q\triangleq \frac{1}{E_{0}}$. It is obvious that, if the first constraint is satisfied $k=1,\ldots,K+1$, then the second constraint is also satisfied. Thus, we can remove the second constraint $k=1,\ldots,K+1$.

The Lagrangian of problem (~\ref{eq8}) is defined as
\begin{equation}
\label{eq9}
\begin{aligned}
\mathcal{L}\Big\{Q,\{A_{k}\}^{K+1}_{k=1},\{\lambda_{k}\}^{K+1}_{k=0}\Big\}&\quad=\frac{1}{Q}+Q\sum^{K+1}_{k=1}\lambda_{k-1}\frac{\sigma^{2}_{k}\beta_{k}\bar{\gamma}_{k}}{\Gamma_{k}}\\&\quad +\sum^{K+1}_{k=1}(\lambda_{k-1}-\lambda_{k})A_{k}-\lambda_{0},
\end{aligned}
\end{equation}
where $\lambda_{k}\geq 0$ and $\lambda_{K+1}\geq 0$ are the dual variables corresponding to the constraints $A_{k-1}-A_{k} \geq Q\frac{\sigma^{2}_{k}\beta_{k}\bar{\gamma}_{k}}{\Gamma_{k}}, k=1,\ldots,K+1$ and $A_{K+1}$, respectively. 
Following (~\ref{eq9}), we can express the KKT conditions as 
\begin{equation}
\label{eq11a}
-\frac{1}{Q^{\star 2}}+\sum^{K+1}_{k=1}\lambda^{\star}_{k-1}\frac{\sigma^{2}_{k}\beta_{k}\bar{\gamma}_{k}}{\Gamma_{k}}=0,
\end{equation}
\begin{equation}
\label{eq12a}
\sum^{K+1}_{k=1}\lambda^{\star}_{k-1}\Bigg(Q^{\star}\frac{\sigma^{2}_{k}\beta_{k}\bar{\gamma}_{k}}{\Gamma_{k}}-(A^{\star}_{k-1}-A^{\star}_{k})\Bigg)=0. 
\end{equation}
If we have $\lambda^{\star}_{k-1}-\lambda^{\star}_{k}>0$, then  the optimal $A^{\star}_{k}$ minimizing the Lagrangian is given by $A^{\star}_{k}=-\infty$, and the dual function is unbounded. Also, if $\lambda^{\star}_{k-1}-\lambda^{\star}_{k}<0$, then we have $A^{\star}_{k}=\infty$, which also yields an unbounded dual function. Therefore, $\lambda^{\star}_{k-1}-\lambda^{\star}_{k}=0, k=1,\ldots,K+1$, this implies that, $\lambda^{\star}_{0}=\lambda^{\star}_{1}=\ldots=\lambda^{\star}_{K+1}$. Hence, the KKT conditions are rewritten as
\begin{equation}
\label{eq11}
-\frac{1}{Q^{\star 2}}+\lambda^{\star}_{0}\sum^{K+1}_{k=1}\frac{\sigma^{2}_{k}\beta_{k}\bar{\gamma}_{k}}{\Gamma_{k}}=0,
\end{equation}
\begin{equation}
\begin{aligned}
\label{eq12}
\lambda^{\star}_{0}\Bigg(Q^{\star}\frac{\sigma^{2}_{k}\beta_{k}\bar{\gamma}_{k}}{\Gamma_{k}}-(A^{\star}_{k-1}-A^{\star}_{k})\Bigg)=0,\text{ } k=1,\ldots,K+1,
\end{aligned}
\end{equation}
\begin{equation}
\begin{aligned}
\label{eq12b}
-\lambda^{\star}_{0}A^{\star}_{K+1}=0.
\end{aligned}
\end{equation}
The optimal minimum source power, $E^{\star}_{0}$, can be acquired from (~\ref{eq11}) as
\begin{equation}
\label{eq14}
E^{\star}_{0}=\frac{1}{Q^{\star}}=\sqrt{\lambda_{0}^{\star}\sum^{K+1}_{k=1}\frac{\sigma^{2}_{k}\beta_{k}\bar{\gamma}_{k}}{\Gamma_{k}}}.
\end{equation}

From the complementary slackness condition for $k=1$ in (~\ref{eq12}), we have the following two cases: 
\begin{equation}
\label{eq15}
\begin{aligned}
&\quad\textit{Case 1:}\text{ }\lambda^{\star}_{0}=0,\text{ and }1-A^{\star}_{1}>Q^{\star}\frac{\sigma^{2}_{1}\beta_{1}\bar{\gamma}_{1}}{\Gamma_{1}},\\&\quad\textit{Case 2:}\text{ }\lambda^{\star}_{0}>0,\text{ and }1-A^{\star}_{1}=Q^{\star}\frac{\sigma^{2}_{1}\beta_{1}\bar{\gamma}_{1}}{\Gamma_{1}}.
\end{aligned}
\end{equation}
For case 1, $\frac{1}{Q}=\sqrt{\lambda^{\star}_{0}\sum^{K+1}_{k=1}\frac{\sigma^{2}_{k}\beta_{k}\bar{\gamma}_{k}}{\Gamma_{k}}}=\infty$, this implies $E_{0}=0$. Obviously, this is not the feasible solution for the problem in equation (~\ref{eq8}) with arbitrary given $\bar{\gamma}_{k}>0$. Thus, we focus on case 2, which yields $\frac{1}{Q}=\sqrt{\lambda^{\star}_{0}\sum^{K+1}_{k=1}\frac{\sigma^{2}_{k}\beta_{k}\bar{\gamma}_{k}}{\Gamma_{k}}}<\infty$, hence $\lambda^{\star}_{0}$ is always positive. Since $\lambda^{\star}_{0}=\lambda^{\star}_{1}=\ldots=\lambda^{\star}_{K+1}$, all the optimal dual variables are positive and equal. From the complementary slackness condition in (~\ref{eq12}), it follows that
\begin{equation}
\label{eq16}
\begin{aligned}
&\quad Q^{\star}\frac{\sigma^{2}_{k+1}\beta_{k+1}\bar{\gamma}_{k+1}}{\Gamma_{k+1}}=(A^{\star}_{k}-A^{\star}_{k+1}), A^{\star}_{K+1}=0,\\&\quad \Bigg(\lambda^{\star}_{0}\sum^{K+1}_{k=1}\frac{\sigma^{2}_{k}\beta_{k}\bar{\gamma}_{k}}{\Gamma_{k}}\Bigg)^{-\frac{1}{2}}\frac{\sigma^{2}_{k+1}\beta_{k+1}\bar{\gamma}_{k+1}}{\Gamma_{k+1}}=(A^{\star}_{k}-A^{\star}_{k+1}),\\&\quad \text{ }k=1,\ldots,K+1.
\end{aligned}
\end{equation}
Since $\sum^{K}_{k=0}(A^{\star}_{k}-A^{\star}_{k+1})=1$, the optimal dual variable $\lambda^{\star}_{0}$ is given by 
\begin{equation}
\label{eq17}
\lambda^{\star}_{0}=\sum^{K+1}_{k=1}\frac{\sigma^{2}_{k}\beta_{k}\bar{\gamma}_{k}}{\Gamma_{k}}, 
\end{equation}
which is obtained from (~\ref{eq12a}). Now, substituting (~\ref{eq17}) into (~\ref{eq14}) yields 
\begin{equation}
\label{eq18}
E^{\star}_{0}=\sum^{K+1}_{k=1}\frac{\sigma^{2}_{k}\beta_{k}\bar{\gamma}_{k}}{\Gamma_{k}}.
\end{equation}
By inserting $E^{\star}_{0}$ into SNR constraint of the optimization problem (~\ref{eq7}), $\rho^{\star}_{k}$ is derived as 
\begin{equation}
\label{eq19}
\rho^{\star}_{k}=\begin{cases}
1-\frac{1}{\prod^{k-1}_{j=1}\rho_{j}}\frac{\frac{\bar{\gamma}_{k}\beta_{k}}{\Gamma_{k}}}{\sum^{K+1}_{j=1}\frac{\bar{\gamma}_{j}\beta_{j}}{\Gamma_{j}}},\text{ }\text{ }\text{  }k=1,\ldots,K \\
0,\text{ }\text{ }\text{  }\text{ }\text{ }\text{ }\text{ }\text{ }\text{ }\text{ }\text{ }\text{ }\text{ }\text{ }\text{ }\text{ }\text{ }\text{ }\text{ }\text{ }\text{ }\text{ }\text{  }k=K+1.
\end{cases}
\end{equation}
The $\rho^{\star}_{k}$ is found by simply making $\rho_{k}$ the subject at equality. $\blacksquare$
\section{Proof of Theorem 2}
\label{app2}
Similar to the proof of Theorem ~\ref{thmoptimal_min}, we apply the change of variables where $A_{k}\triangleq\prod^{k}_{j=1}\rho_{j}$, $A_{0}\triangleq 1$ and $A_{K+1}\triangleq 0$, since problem (~\ref{eq22}) is also non-convex with respect to $\hat{\gamma}$ and $\{\rho_{k}\}^{K}_{k=1}$. Then, the non-convex problem in (~\ref{eq22}) can be equivalently transformed to the following convex formulation
\begin{equation}
\label{eq23}
\begin{aligned}
& \underset{\{A_{k}\}^{K+1}_{k=1}}{\text{maximize}}\text{ }\hat{\gamma}\\
& \text{subject to} 
\begin{aligned}  
& & & A_{k-1}-A_{k} \geq \hat{\gamma}\frac{\sigma^{2}_{k}\beta_{k}}{E_{0}\Gamma_{k}}, \text{ }k=1,\ldots,K+1,
\end{aligned}\\
& \text{ } 
\begin{aligned}  
& & & & & & & & & & A_{k} \leq A_{k-1}, \text{ }k=1,\ldots,K+1.
\end{aligned}
\end{aligned}
\end{equation}

By solving the first, we solve the second constraint of problem (~\ref{eq23}), therefore, the second constraint is removed (i.e., it is not considered in the Lagrangian and solution). The Lagrangian for problem (~\ref{eq23}) is thus expressed as
\begin{equation}
\label{eq24}
\begin{aligned}
\mathcal{L}\Big\{\hat{\gamma},\{A_{k}\}^{K+1}_{k=1},\{\lambda_{k}\}^{K+1}_{k=0}\Big\}&\quad= \hat{\gamma}+\hat{\gamma}\sum^{K+1}_{k=1}\lambda_{k-1}\frac{\sigma^{2}_{k}\beta_{k}}{E_{0}\Gamma_{k}}\\&\quad+\sum^{K+1}_{k=1}(\lambda_{k-1}-\lambda_{k})A_{k}-\lambda_{0},
\end{aligned}
\end{equation}
with its KKT conditions given as 
\begin{equation}
\label{eq25}
1+\sum^{K+1}_{k=1}\lambda^{\star}_{k-1}\frac{\sigma^{2}_{k}\beta_{k}}{E_{0}\Gamma_{k}}=0,
\end{equation}
\begin{equation}
\label{eq26}
\lambda^{\star}_{k-1}-\lambda^{\star}_{k}=0,
\end{equation}
and
\begin{equation}
\label{eq27}
\sum^{K+1}_{k=1}\lambda^{\star}_{k-1}\Bigg(\hat{\gamma}^{\star}\frac{\sigma^{2}_{k}\beta_{k}}{E_{0}\Gamma_{k}}-(A^{\star}_{k-1}-A^{\star}_{k})\Bigg)=0.
\end{equation}

From (~\ref{eq26}), we deduce that $\lambda^{\star}_{k-1}=\lambda^{\star}_{k}$, hence, $\lambda^{\star}_{0}=\lambda^{\star}_{1}=\ldots=\lambda^{\star}_{K+1}$. The KKT conditions are modified to give,
\begin{equation}
\label{eq28}
1+\lambda^{\star}_{0}\sum^{K+1}_{k=1}\frac{\sigma^{2}_{k}\beta_{k}}{E_{0}\Gamma_{k}}=0,
\end{equation}
\begin{equation}
\label{eq29}
\lambda^{\star}_{0}\Bigg(\hat{\gamma}^{\star}\frac{\sigma^{2}_{k}\beta_{k}}{E_{0}\Gamma_{k}}-(A^{\star}_{k-1}-A^{\star}_{k})\Bigg)=0,\text{ }
-\lambda^{\star}_{0}A^{\star}_{K+1}=0.
\end{equation}
The optimal $\lambda^{\star}_{0}$ can be derived from (~\ref{eq28}) as
\begin{equation}
\label{eq31}
\lambda^{\star}_{0}=-\frac{1}{\sum^{K+1}_{k=1}\frac{\sigma^{2}_{k}\beta_{k}\hat{\gamma}^{\star}}{E_{0}\Gamma_{k}}}=-\frac{E_{0}}{\sum^{K+1}_{k=1}\frac{\sigma^{2}_{k}\beta_{k}\hat{\gamma}^{\star}}{\Gamma_{k}}}.
\end{equation}
Noting that $\lambda^{\star}_{0}=\lambda^{\star}_{1}=\ldots=\lambda^{\star}_{K+1}$, the optimal dual variables are negative and equal based on (~\ref{eq31}). To find the optimal $\hat{\gamma}^{\star}$, (~\ref{eq31}) is substituted into (~\ref{eq29}) to produce ,
\begin{equation}
\begin{aligned}
\label{eq33}
&\quad\frac{E_{0}}{\sum^{K+1}_{k=1}\frac{\sigma^{2}_{k}\beta_{k}\hat{\gamma}^{\star}}{\Gamma_{k}}}\sum^{K+1}_{k=1}(A^{\star}_{k-1}-A^{\star}_{k})=\Bigg(\frac{\sum^{K+1}_{k=1}\frac{\hat{\gamma}^{\star}\sigma^{2}_{k}\beta_{k}E_{0}}{E_{0}\Gamma_{k}}}{\sum^{K+1}_{k=1}\frac{\sigma^{2}_{k}\beta_{k}\hat{\gamma}^{\star}}{\Gamma_{k}}}\Bigg)\\&\quad=1.
\end{aligned}
\end{equation} 
Since $\sum^{K+1}_{k=1}(A^{\star}_{k-1}-A^{\star}_{k})=1$, the optimal $\hat{\gamma}^{\star}$ is obtained as 
\begin{equation}
\label{eq32}
\hat{\gamma}^{\star}=\frac{E_{0}}{\sum^{K+1}_{k=1}\frac{\sigma^{2}_{k}\beta_{k}}{\Gamma_{k}}}.
\end{equation}
By placing $\hat{\gamma}^{\star}$ into the SNR constraint of problem (~\ref{eq23}), making $\rho_{k}$ the subject at equality, the optimal $\rho^{\star}_{k}$ can be found as 
\begin{equation}
\label{eq34}
\rho^{\star}_{k}=\begin{cases}
1-\frac{1}{\prod^{k-1}_{j=1}\rho_{j}}\frac{\frac{\beta_{k}}{\Gamma_{k}}}{\sum^{K+1}_{j=1}\frac{\beta_{j}}{\Gamma_{j}}},\text{ }\text{ }k=1,\ldots,K \\
0,\text{ }\text{ }\text{  }\text{ }\text{ }\text{ }\text{ }\text{ }\text{ }\text{ }\text{ }\text{ }\text{ }\text{ }\text{ }\text{ }\text{ }\text{ }\text{ }\text{ }\text{ }\text{ }k=K+1.
\end{cases}
\end{equation}

\begin{figure}[t!]
\begin{center}
\includegraphics[width=3.4in]{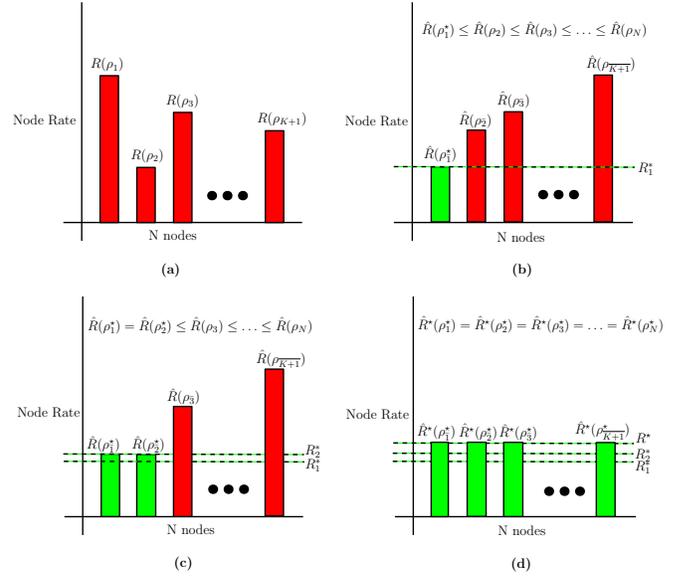}
\end{center}
\caption{A graphical explanation of how the optimal system rate is achieved: (a) shows the achievable rate by each of the DF-SWIPT relay nodes and the destination node, (b) the node rates are rearranged in ascending order and the minimum rate's PS ratio is optimized, (c) the first and second node adjust their individual rates based on their PS ratio until they achieve the same rate,(d) each successive node's rate is adjusted based on their PS ratio until the whole system achieves the optimal system rate.}
\label{figproofmin}
\end{figure}

We will now show that $\hat{\gamma}^{\star}$ is the optimal SNR threshold with the aid of both simple arithmetic and the diagrams presented in Fig. ~\ref{figproofmin}. We know that each of the $K+1$ nodes (i.e., the DF-SWIPT relay nodes and the destination node) achieve different individual rates, \{$R(\rho_{k})\}^{K+1}_{k=1}$ depending on their current $\rho_{k}$ value as shown in Fig. ~\ref{figproofmin}(a). These rates can be sorted in an ascending order as $\hat{R}(\rho_{\bar{1}})\leq\hat{R}(\rho_{\bar{2}})\leq\ldots\leq\hat{R}(\rho_{\overline{K+1}})$, where $\bar{k}=\bar{1},\bar{2},\ldots,\overline{K+1}$ is the new ordering number for each of the $K+1$ nodes based on the achieved rates. Also, its obvious that a node's rate is a function of its $\rho_{\bar{k}}$. A node's rate increases or decreases by reducing or increasing its $\rho_{\bar{k}}$, respectively. The system throughput at this first stage is constrained by $\hat{R}(\rho_{\bar{1}})$ since it is the minimum achievable rate as seen in Fig. ~\ref{figproofmin}(b). In Fig. ~\ref{figproofmin}(c), by reducing $\rho_{\bar{1}}$ and increasing $\rho_{\bar{2}}$, the system can achieve a state where $\hat{R}(\rho_{\bar{1}})=\hat{R}(\rho_{\bar{2}})\leq\ldots\leq\hat{R}(\rho_{\overline{K+1}})$. Now the system throughput is constrained by the minimum rate $\hat{R}(\rho_{\bar{1}})=\hat{R}(\rho_{\bar{2}})$. Following the same procedure for the preceding node, the system achieves a rate of $\hat{R}(\rho_{\bar{1}})=\hat{R}(\rho_{\bar{2}})=\hat{R}^(\rho_{\bar{3}})\leq\ldots\leq\hat{R}(\rho_{\overline{K+1}})$. By continuously repeating and applying this same process to all the subsequent nodes, the system will achieve a rate where $\hat{R}^{\star}(\rho^{\star}_{\bar{1}})=\hat{R}^{\star}(\rho^{\star}_{\bar{2}})=\ldots=\hat{R}^{\star}(\rho^{\star}_{\overline{K+1}})$ as shown in Fig. ~\ref{figproofmin}(d). From our solution, we are able to acquire closed-form solutions for our PS ratio at each of the $K+1$ nodes. These optimal PS ratios achieve the maximized set system SNR (i.e., $\{\hat{\gamma}\}^{K+1}_{k=1}=\hat{\gamma}^{\star}$), this implies that, each node attains the maximized system rate established as $R^{\star}=\log_{2}(1+\hat{\gamma}^{\star})$. $\blacksquare$
\section{Number of Relay Nodes Estimation}
\label{app3}
From (~\ref{eqgeneraleq}), we have the estimated source power as 
\begin{equation}
\label{app11}
\begin{aligned}
E_{0}&\quad\approx \sum^{K+1}_{k=1}\frac{\bar{\gamma}\sigma^{2}_{k}\beta_{k}}{\mathop{\mathbb{E}}[\Gamma_{k}]}=  \sum^{K+1}_{k=1}\frac{\bar{\gamma}\sigma^{2}_{k}\beta_{k}}{\mathop{\mathbb{E}}\Big[\prod^{k}_{j=1}C_{j}\Big(\frac{d_{j}}{d_{0}}\Big)^{-\alpha_{j}}\beta_{j}\vert h_{j}\vert^{2}\Big]}\\&\quad = \sum^{K+1}_{k=1}\frac{\bar{\gamma}\sigma^{2}_{k}\beta_{k}}{\prod^{k}_{j=1}C_{j}\Big(\frac{d_{j}}{d_{0}}\Big)^{-\alpha_{j}}\beta_{j}\mathop{\mathbb{E}}\Big[\prod^{k}_{j=1}\vert h_{j}\vert^{2}\Big]}\\&\quad = \sum^{K+1}_{k=1}\frac{\bar{\gamma}\sigma^{2}_{k}\beta_{k}}{\prod^{k}_{j=1}C_{j}\Big(\frac{d_{j}}{d_{0}}\Big)^{-\alpha_{j}}\beta_{j}\prod^{k}_{j=1}\mathop{\mathbb{E}}\big[\vert h_{j}\vert^{2}\big]}.
\end{aligned}
\end{equation}
Assuming the sensor network is a homogeneous sensor network, let $\beta_{1}=\beta_{2}=\ldots=\beta_{K}$, $\beta_{K+1}=1$, $\sigma_{1}=\sigma_{2}=\ldots=\sigma_{K}=\sigma_{K+1}$, and $d_{1}=d_{2}=\ldots=d_{K}=d_{K+1}$, then, 
\begin{equation}
\label{app12}
\begin{aligned}
E_{0}&\quad\approx \sum^{K+1}_{k=1}\frac{\bar{\gamma}\sigma^{2}\beta}{\mathop{C^{k}\Big(\frac{d}{d_{0}}\Big)^{-k\alpha_{j}}\beta^{k}\prod^{k}_{j=1}\mathbb{E}}\big[\vert h_{j}\vert^{2}\big]}.
\end{aligned}
\end{equation}
Since $\vert h_{j}\vert^{2}$ is a random variable with an exponential distribution, hence, $\mathop{\mathbb{E}}\big[\vert h_{j}\vert^{2}\big]=1/2=0.5$, because $h_{j} \sim \mathcal{CN}(0,1)$. This expectation solution is acquired from the fact that, the expectation of an exponential distribution acquired from a the square of Rayleigh distribution is the variance of the Rayleigh distribution divided by $2$, hence, we have $1/2$. Therefore the optimal source power becomes
\begin{equation}
\label{app12a}
\begin{aligned}
E_{0}&\quad\approx \sum^{K+1}_{k=1}\frac{\bar{\gamma}\sigma^{2}\beta}{C^{k}\Big(\frac{d}{d_{0}}\Big)^{-k\alpha_{j}}\beta^{k}0.5^{k}}\\&\quad= \sum^{K+1}_{k=1}\frac{\bar{\gamma}\sigma^{2}\beta}{\Big[ 0.5 C\Big(\frac{d}{d_{0}}\Big)^{-\alpha}\beta\Big]^{k}}.
\end{aligned}
\end{equation}
Let $\Gamma =0.5 C\Big(\frac{d}{d_{0}}\Big)^{-\alpha}\beta$, then, the approximated source power becomes, 
\begin{equation}
\label{app12b}
\begin{aligned}
E_{0}\approx \sum^{K+1}_{k=1}\frac{\bar{\gamma}\sigma^{2}\beta}{\Gamma^{k}}.
\end{aligned}
\end{equation}
Using the approximation $E_{0}\approx \sum^{N}_{k=1}\frac{\bar{\gamma}\sigma^{2}\beta}{\Gamma^{k}}$, we present the stepwise derivation of the estimated maximum number of DF-SWIPT relay nodes supported by a given source power. By expanding the (~\ref{app12b}), we have, 
\begin{equation}
\label{app13}
\begin{aligned}
E_{0}&\quad\approx \sum^{K+1}_{k=1}\frac{\bar{\gamma}\sigma^{2}\beta}{\Gamma^{k}}= \sum^{K+1}_{k=0}\frac{\bar{\gamma}\sigma^{2}\beta}{\Gamma^{k}}-\bar{\gamma}\sigma^{2}\beta\\&\quad=\frac{\bar{\gamma}\sigma^{2}\beta}{\Gamma^{0}}+\frac{\bar{\gamma}\sigma^{2}\beta}{\Gamma^{1}}+\frac{\bar{\gamma}\sigma^{2}\beta}{\Gamma^{2}}+\ldots+\frac{\bar{\gamma}\sigma^{2}\beta}{\Gamma^{K+1}}-\bar{\gamma}\sigma^{2}\beta.
\end{aligned}
\end{equation}
It is obvious from (~\ref{app11}) that the $E_{0}$ equation is similar to the geometric progression sum equation, that is,
\begin{equation}
\label{app14}
\begin{aligned}
E_{0}&\quad\approx \alpha_{0}r^{0}+\alpha_{0}r^{1}+\alpha_{0}r^{2}+\ldots+\alpha_{0}r^{K+1}-\alpha_{0}\\&\quad = \alpha_{0}\Big(\frac{1-r^{K+1}}{1-r}\Big)-\alpha_{0}=\alpha_{0}\Big(\frac{1-r^{K+1}}{1-r}-1\Big),
\end{aligned}
\end{equation}
where $\alpha_{0}=\bar{\gamma}\sigma^{2}\beta$ and $r=\frac{1}{\Gamma}$ ~\cite{Bird14,Stroud13}. By making $K+1$ the subject, we have,
\begin{equation}
\label{app15}
E_{0}\approx\alpha_{0}\Big(\frac{1-r^{K+1}}{1-r}-1\Big),
\end{equation}
\begin{equation}
\label{app16}
1-\Big(\frac{E_{0}}{\alpha_{0}}+1\Big)\Big(1-r\Big)\approx r^{K+1},
\end{equation}
\begin{equation}
\label{app17}
\ln\Big[1-\Big(\frac{E_{0}}{\alpha_{0}}+1\Big)\Big(1-r\Big)\Big]\approx(K+1)\ln r,
\end{equation}
\begin{equation}
\label{app18}
K+1\approx\frac{\ln\Big[1-\Big(\frac{E_{0}}{\alpha_{0}}+1\Big)\Big(1-r\Big)\Big]}{\ln r}.
\end{equation}
The above solution is acquired from the geometric progression sum equation is for when $0 <r < 1$, that is, $\Gamma > 1$. Following similar steps used above, when $r > 1$, that is, $\Gamma < 1$, the solution becomes,
\begin{equation}
\label{app19}
K+1\approx\frac{\ln\Big[1+\Big(\frac{E_{0}}{\alpha_{0}}+1\Big)\Big(r-1\Big)\Big]}{\ln r}.
\end{equation}
Now, we replace $\alpha_{0}$ and $r$ by their defined values to acquire the approximation of $K$ DF-SWIPT nodes as
\begin{equation}
\label{appnodenumb}
K\approx\begin{cases}
\frac{\ln\Big[1-\Big(\frac{E_{0}}{\bar{\gamma}\sigma^{2}\beta}+1\Big)\Big(1-\frac{1}{\Gamma}\Big)\Big]}{-\ln \Gamma}-1,\text{ }\text{ }0<r<1,\Gamma > 1 \\
\frac{\ln\Big[1+\Big(\frac{E_{0}}{\bar{\gamma}\sigma^{2}\beta}+1\Big)\Big(\frac{1}{\Gamma}-1\Big)\Big]}{-\ln \Gamma}-1,\text{ }\text{ }r>1,\Gamma <1.
\end{cases}
\end{equation}
It can be observed that equation ($~\ref{appnodenumb}$) is always positive and depends on the value of $\Gamma$. $\blacksquare$
\ifCLASSOPTIONcaptionsoff
  \newpage
\fi
\bibliographystyle{IEEEtr}

\end{document}